\documentstyle[11pt,newpasp,twoside,epsf]{article}
\def\spose#1{\hbox to 0pt{#1\hss}}
\def\simlt{\mathrel{\spose{\lower 3pt\hbox{$\mathchar"218$}}
     \raise 2.0pt\hbox{$\mathchar"13C$}}}
\def\simgt{\mathrel{\spose{\lower 3pt\hbox{$\mathchar"218$}}
     \raise 2.0pt\hbox{$\mathchar"13E$}}}
\def\edcomment#1{\iffalse\marginpar{\raggedright\sl#1\/}\else\relax\fi}
\reversemarginpar

\begin{document}
\title{Are High-Velocity Clouds the Building Blocks of the Local Group?}
\author{Brad K. Gibson, Yeshe Fenner, Sarah T. Maddison, Daisuke Kawata}
\affil{Centre for Astrophysics \& Supercomputing, Swinburne University,
Mail \#31, P.O. Box 218, Hawthorn, Victoria, 3122  Australia}

\begin{abstract}
Motivated by the apparent order-of-magnitude discrepancy
between the observed number of Local Group satellite galaxies, and
that predicted by $\Lambda$CDM hierarchical clustering cosmologies,
we explore an alternate suggestion - perhaps the missing satellites are 
not actually ``missing'', but are instead ``in disguise''.  The disguise
we consider here is that of the classical H{\small I} High-Velocity Clouds.
Is it possible that what have been thought of traditionally as
a ``Galactic'' phenomenon, are actually the building blocks of the 
Local Group?  We discuss the strengths and weaknesses of this 
hypothesis, and highlight avenues of future research
which may provide an unequivocal resolution to this contentious issue.
\end{abstract}

\vspace{-5.0mm}
\section{Introduction}
A natural byproduct of hierarchical clustering galaxy formation 
scenarios - perhaps best represented by the currently favoured 
$\Lambda$-dominated Cold Dark Matter ($\Lambda$CDM) paradigm - is that our
Local Group of galaxies should be populated by 
$\sim$500 satellite objects (e.g. Klypin et~al. 1999).
This prediction is in (apparent) stark contradiction with the observed Local
Group census ($\sim$30 satellites).

While the above dichotomy makes it tempting to pursue 
alternatives to $\Lambda$CDM, it would be
prudent to at least consider the more conservative hypothesis - 
$\Lambda$CDM is correct and we have simply ``misplaced'' the satellites.
Two obvious mechanisms consistent with this hypothesis would be to\hfil\break
\indent
1. make the satellites invisible, or\hfil\break
\indent
2. disguise the satellites.\hfil\break
\noindent
Options under mechanism 1 include stripping/ejecting any baryons
associated with the CDM halo, perhaps through feedback
(Chiu, Gnedin \& Ostriker 2001),
and/or ionising any residual gas therein, or perhaps through re-ionisation
of the Universe not allowing gas to cool into the smallest CDM
halos (Moore 2001).
In both cases, little accompanying stellar component can exist
(or else the satellites would become ``visible'').
In some sense, mechanism 2 represents the least ``radical'' of the options
and forms the basis of much of the subsequent discussion.

To date, the best suggested ``disguise'' for these supposed missing
satellites is that worn by the population of High-Velocity Clouds (HVCs).
HVCs are traditionally classified as H{\small I} gas clouds whose velocities
are inconsistent with that of Galactic rotation (e.g. Wakker, van Woerden \&
Gibson 1999).  Since their discovery
40 years ago, debate concerning their origin has ranged from
the local (Galactic fountain), to the
intermediate (tidal disruption of accreting
dwarfs), to the distant ($\Lambda$CDM building blocks
left over from the formation of the Local Group).
Blitz et~al. (1999) and later Braun \& Burton (1999), revived this
extragalactic scenario,\footnote{In contrast with Blitz et~al. (1999), 
Braun \& Burton (1999) eliminate all extended ($\simgt$1$^\circ$) HVCs, and
consider only Compact HVCs (CHVCs) to be associated with
residual $\Lambda$CDM halos.}
and have both provided persuasive arguments
to support their cases (see also Blitz 2001; Burton et~al. 2001).
With $\sim$2000 HVCs now catalogued (Putman et~al. 2001), their
numbers are certainly a good match to the predicted number of $\Lambda$CDM
halos.  Shortcomings to this picture have been highlighted 
by Gibson et~al. (2001ab), Weiner et~al. (2001),
Zwaan (2001), Charlton et~al. (2000), and Combes \&
Charmandaris (2000), amongst others.

In what follows, we discuss several arguments
which have been employed to support (or refute)
the extragalactic HVC scenario,
drawing attention to ongoing (and future)
programs designed to shed light on this contentious issue.

\section{The Evidence ...}
In terms of discriminating between the extragalactic ($\Lambda$CDM)
and Galactic (fountain or dwarf galaxy tidal disruption) origin scenarios 
for HVCs, any number of \it indirect \rm arguments can be made,
but in actuality, only one, clean, \it direct \rm discriminant exists - 
HVC distance.

\subsection{Distances}
Under the $\Lambda$CDM scenario, HVCs are presumed to populate
the Local Group, with typical distances of 
$\sim$700\,kpc; in contrast, 
both the Galactic fountain and dwarf galaxy disruption scenarios favour
distances of order $\sim$10\,kpc.  In theory,
this (approximate) two orders-of-magnitude difference offers a clean
discriminant between the models.  In practice, despite distance 
being this ``Holy Grail'', its determination is \it extremely \rm challenging.

The only bona fide mechanism for setting a useful HVC
distance bracket (or an upper limit) is via absorption line spectroscopy
towards background halo stars of known distance.  As stressed by Wakker
(2001) and Gibson et~al. (2001a), the dearth of suitably
bright and distant, blue-horizontal branch halo stars aligned with the high
H{\small I} column density cores of HVCs,\footnote{We are undertaking a
program at the United Kingdom Schmidt Telescope, in an attempt to rectify this
shortcoming.  We (in collaboration
with Mike Bessell, Tim Beers, Norbert Christlieb, John Norris,
and Joss Bland-Hawthorn) are employing the 
6dF facility to take intermediate-resolution 
spectra of Hamburg/ESO Survey and Beers/Preston HK Survey
candidate B-HB halo stars, prior to followup
high-resolution work for the best HVC probes.} has limited the
successful\footnote{``Success'' defined here as a clear halo vs
non-halo residency determination
for the given HVC.} application of this
technique to just five HVCs.\footnote{The marginal
detection of an HVC in Complex~WD, seen in absorption against a
halo RR~Lyrae at 5\,kpc (Comeron 2000, priv comm), 
needs confirmation, 
particularly in light of its projected location near the Local Group's
anti-barycentre.}  \bf All five of these HVCs lie clearly in the Galactic
halo\rm\footnote{Despite Blitz's (2001) protestations, HVC~100$-$7$+$100
is both an HVC - its positive velocity is significantly inconsistent
with that expected by Galactic rotation in this quadrant - and its existence
confirmed in both HI (Bates et~al. 1991) and the ultraviolet
(Bates et~al. 1990).  Its low column density - N(HI)=
3$\times$10$^{18}$\,cm$^{-2}$ - is what makes it difficult to extract from
the Leiden-Dwingeloo Survey data, and not its non-existence!}
Gibson et~al. (2001a; Table~1).

While it might be tempting to succomb to hyperbole and claim the HVC mystery
solved, since all five for which a halo vs non-halo status could be
determined clearly reside in the Galactic halo, it's crucial to bear in mind
that (i) we are still only talking about $\sim$1\% of the known HVCs, and (ii)
an unavoidable selection effect plagues this interpretation, in the sense that 
there simply aren't any background halo stars out at $\simgt$500\,kpc with
which to probe HVCs at (putative) comparable distances.
\it If \rm it could be shown that gaseous HVCs are accompanied by an
associated stellar population, it is not inconceivable that deep, targeted,
searches for the tip of the red giant branch (or maybe RR~Lyrae) 
might provide a useful distance determination for intra-Local Group
HVCs.  Using the LCO and KPNO, Grebel et~al. (2000) are searching
for RGB stars in several CHVCs, and while they have several
candidates, the data do not show convincing evidence for the presence of
significant numbers of stars.  Unequivocal conclusions are complicated by the
potential confusion with unresolved, background, starburst galaxies.  8-10m
class spectroscopic confirmation for all candidates is a necessity.
Josh Simon, as part of his PhD at Berkeley,
has also been cross-correlating hundreds of HVCs with POSS-II 
plates.\footnote{We are involved in several deep 
searches for stars in CHVCs, involving the 6.5m Baade Telescope and 
(hopefully!) the 4m Anglo-Australian Telescope.  Those involved in the 
collaborations include Dan Kelson, Wendy Freedman, Geraint Lewis, 
Rodrigo Ibata, and Joss Bland-Hawthorn.}

While the absorption line and tip of the RGB
techniques are classified as {\it direct}, they are
applicable to only a handful of HVCs.  That said, there are several
{\it indirect} methods which can be applied to (potentially) hundreds of
HVCs.  

The indirect technique 
which has received the most attention over the past few years
is that based upon H$\alpha$ emission (Bland-Hawthorn \& Maloney 1999).
As shown by Bland-Hawthorn et~al. (1998),
Tufte et~al. (1998), and Weiner et~al. (2001), HVCs are detected regularly in
H$\alpha$ emission, with a subset seen in low ionisation lines.  Under the
(reasonable) 
assumption that $\simgt$1\% of the Galaxy's ionising photons escape the disk,
coupled with models governing the distribution of this ionising field, 
covering fraction, topology, and line-of-sight orientation, an H{\small I}
screen at a given distance will result in a specific H$\alpha$ emission
measure.  The original halo radiation field of Bland-Hawthorn \& Maloney
assumed an underlying exponential disk for the Galaxy.
While valid in the far-field limit, this model broke down within
$\sim$10\,kpc of the disk, where the proximity to spiral 
arms is important (and where we know many HVCs lie).  However,
as reported at this meeting, most failings of the preliminary
model have been rectified by incorporating a proper treatment
of spiral arms (Bland-Hawthorn et~al. 2001).
All known HVCs, with both measured
H$\alpha$ emission and absorption line distance constraints, are now
consistent with the predictions of this 
spiral arm H$\alpha$ model, and appear to lie 
within $\sim$100\,kpc of the Galaxy; to date, this indirect 
technique does \it not \rm support the intra-Local Group
predictions of Blitz et~al. (1999) and Braun \& Burton (1999).
What remains true however, is that H$\alpha$
detections of the Magellanic Stream are perplexingly bright,
and require at least one additional (unidentified) ionising source
(Weiner et~al. 2001; Bland-Hawthorn et~al. 2001).

Employing thermal pressure arguments pertaining to
CHVCs,\footnote{Specifically, the physical conditions necessary for the
establishment (and maintenance) of cold neutral cores embedded within
warm neutral halos.} Burton et~al. (2001) claim that CHVCs lie at a 
distance of 400$\pm$280\,kpc.  As emphasised to us
by Amiel Sternberg (2001, priv comm)\footnote{Details are provided in their
forthcoming paper (Sternberg et~al. 2001).} though, distance
estimates based upon eqn~4 of Burton et~al. are technically only
upper limits.  Their argument is predicated upon the
assumption that the thermal pressure (P) equates to the minimum pressure
(P$_{\rm min}$) required at the core/halo interface.  However, the
condition for a multiphased mixture is that P$_{\rm min}$$<$P$<$P$_{\rm max}$;
P could be significantly larger than P$_{\rm min}$ and still be
consistent with cold and warm neutral media (provided, of course, that
P$<$P$_{\rm max}$).  Even these upper limits are
uncertain, since the value of P$_{\rm min}$ depends upon the assumed 
shielding column and there is no \it a priori \rm reason to
adopt 1$\times$10$^{19}$\,cm$^{-2}$ (as was used by Burton et~al.).

HVC~165$-$43$-$120 is part of the Anti-Centre High-Velocity complex
mapped in H{\small I} by Cohen (1981), and H$\alpha$ by Weiner et~al. (2001).
The Cohen ``Stream'' at $-$110\,km/s spans 25$^\circ$ on the sky and
traces a parallel filament clearly associated with the local ISM
at $-$13\,km/s.  Cohen suggests the $-$110\,km/s must therefore be colliding
with the Galactic disk at a distance of $\sim$300\,pc.
This Stream is akin to the family of HVCs which show direct connections to
Galactic gas (Putman \& Gibson 1999), and therefore must be relatively
nearby.

Building upon the observation that a large number ($\sim$20\%)
of CHVCs show compression-front and tail-shaped features, suggestive of
interaction with an external medium, Quilis \& Moore (2001) have performed
3D-hydro simulations which are consistent with the data, but only
for ambient densities $\simgt$10$^{-4}$\,cm$^{-3}$.  Such densities are
not consistent with that expected in the intergalactic medium.
These simulations suggest that $\simgt$20\% of CHVCs reside in the halo of
our Galaxy.

In summary, those HVCs which possess useful \it direct \rm
distance constraints all reside within the halo of the Milky Way.  Similarly,
\it indirect \rm distance measurements based upon H$\alpha$ emission, 
relation to Galactic disk H{\small I}, and head-tail substructure, are 
also consistent with a halo residency.

\subsection{Metallicities}

While HVC distance is the cleanest discriminant between Galactic \it 
fuel \rm and galactic \it waste \rm scenarios, metallicity also offers
a potentially useful, albeit indirect, indicator.  Under the 
Blitz et~al. (1999) and Braun \& Burton (1999) scenarios, HVCs (or at 
least CHVCs) can be considered to be remnants (or building blocks, 
or perhaps ``failed'' dwarfs) of the formation of the Local Group.
Blitz et~al. suggest that metallicities $\simlt$0.2\,Z$_\odot$ are consistent
with their model.
If HVCs really are failed dwarfs, though (or accreting remnants
of the early phases of the Local Group), one might expect their metallicities
to be $\simlt$0.01\,Z$_\odot$, since such metallicities are encountered in the
dwarf spheroidals (dSphs) of the Local Group.  However,
as summarised by Gibson et~al. (2001a), 
the majority of HVCs appear to have
metallicities of $\sim$0.3\,Z$_\odot$, subject to the usual
caveats concerning ionisation and dust corrections, as well as small-scale 
H{\small I} sub-structure.

Inspection of Figure~1 illustrates dramatically the discrepancy between
HVC metallicities (upper boxed region) and metallicities of the
lowest luminosity components of the Local Group (lower boxed
region).  If HVCs really are Local Group building blocks,
they are clearly very different from the dSphs we
currently see.

\begin{figure}[ht]
\plotfiddle{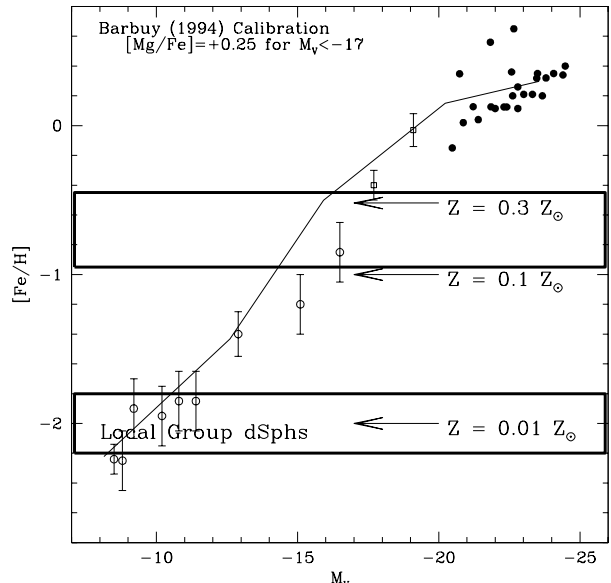}{2.70in}{0}{42}{42}{-135}{-80}
\caption{\footnotesize
Spheroidal galaxy metallicity-luminosity relation, adapted from
Gibson (1997).  Local Group dwarf spheroidals, akin perhaps to the Group's
original building blocks, show metallicities of $\sim$0.01\,Z$_\odot$ 
(lower boxed region), while HVCs show values in the range 
$\sim$0.1$\rightarrow$0.3\,Z$_\odot$.  \bf HVCs today do not resemble
Local Group building blocks such as low luminosity dwarf galaxies.\rm
}
\end{figure}

HVC Complex~C deserves a few words, as it has been put forth as the most
likely candidate to be the low-metallicity infalling Galactic fuel required
by chemical evolution models aiming to avoid the so-called ``G-dwarf
problem''.  This problem is the (generally) unavoidable 
overproduction of low-metallicity stars in closed-box models of the
Galaxy's evolution.  Wakker et~al. (1999) derived a metallicity of
0.09\,Z$_\odot$ based upon the Mrk~290 sightline through Complex~C, and
suggested this HVC was this infalling Galactic star formation fuel.
This interpretation has since been clouded by the analysis of
Gibson et~al. (2001b), who show that abundances as high as 
$\sim$0.3\,Z$_\odot$ are encountered along the Mrk~817 
sightline.  As noted by Tosi (1988), infalling gas metallicities
$\simgt$0.2\,Z$_\odot$ lead to a present-day disk
gas-phase oxygen gradient which is inconsistent with that observed.
We have returned to this issue with our new dual-infall
model for Galactic chemical evolution, employing {\tt GEtool}, a new Galaxy
Evolution tool under development at Swinburne
(Fenner \& Gibson 2002).  Our results strengthen the conclusions of Tosi,
not only through the oxygen gradients, but also the G-dwarf distribution, 
age-metallicity relation, and gas surface density constraints.
Preliminary results are shown in Figure~2, with full details to be 
provided in Chiappini et~al. (2002).

\newpage

\begin{figure}[ht]
\plotfiddle{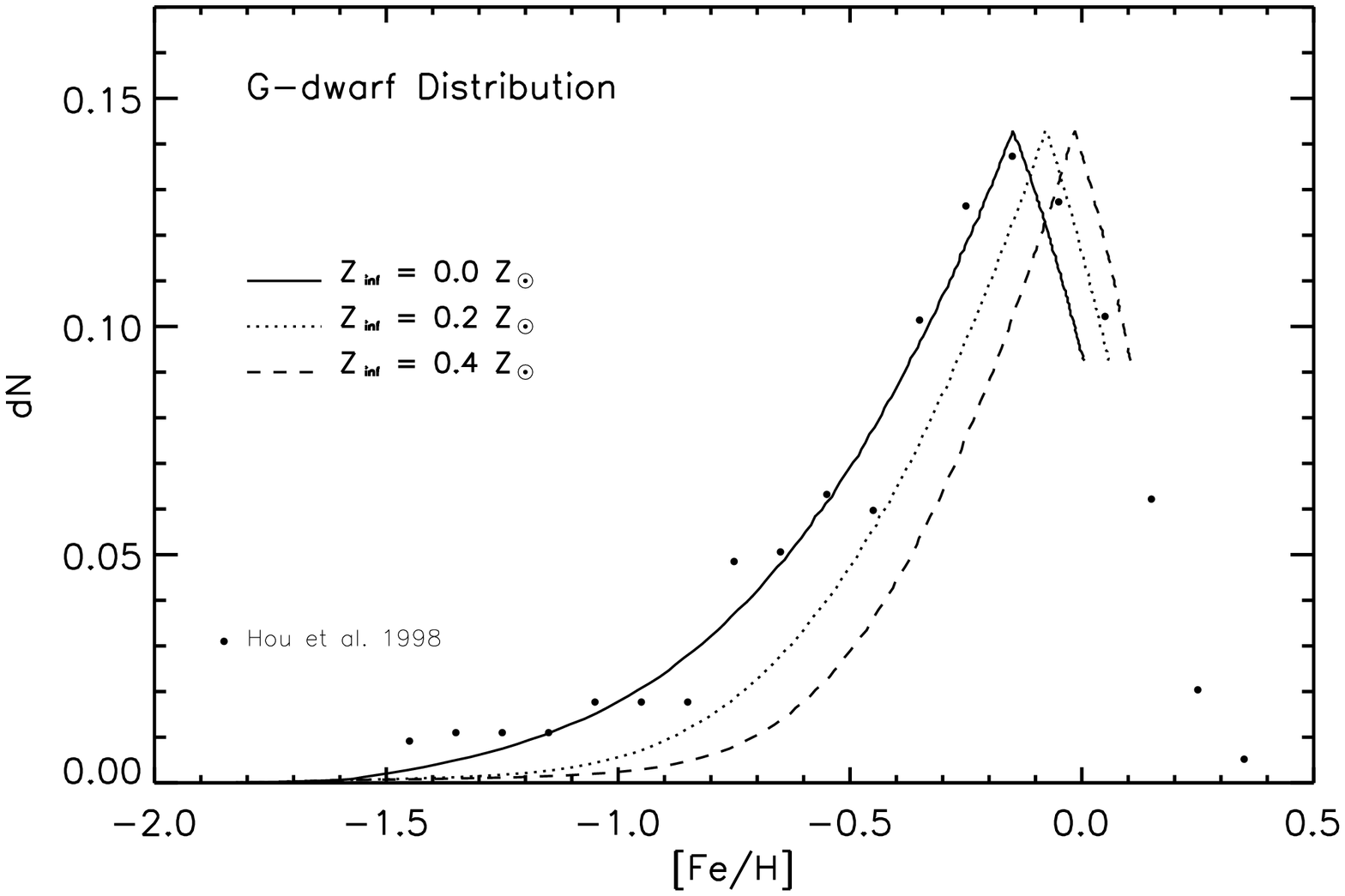}{2.00in}{0}{41}{41}{-140}{-10}
\plotfiddle{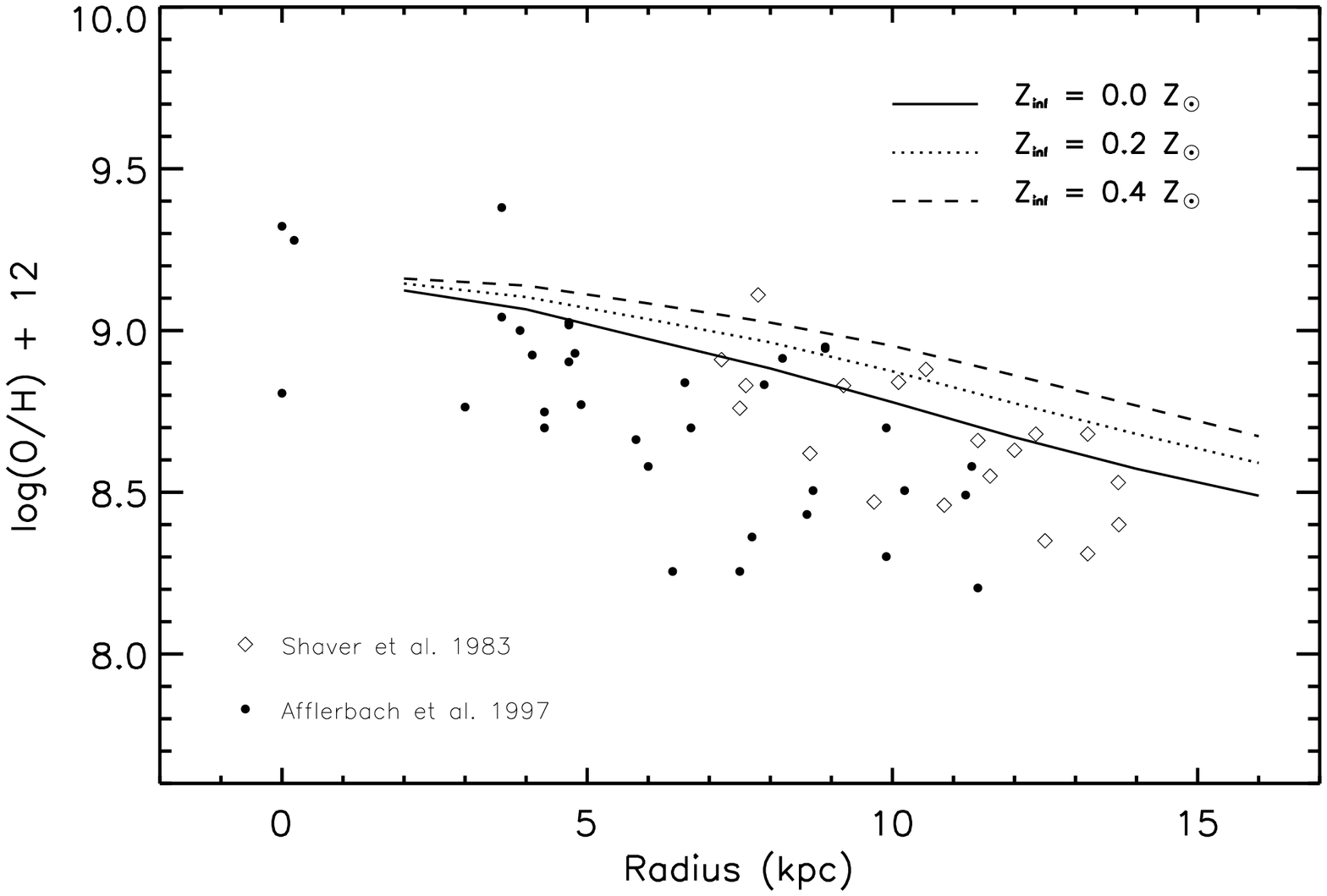}{2.00in}{0}{41}{41}{-140}{-15}
\caption{\footnotesize\it Upper: 
\rm Solar neighbourhood G-dwarf distribution - symbols
represent the observed data, while curves correspond to model predictions
from {\tt GEtool} (Fenner \& Gibson 2002), each of which employed a 
different metallicity for the infalling gas (Z$_{\rm inf}$)
in the disk-phase of our dual-infall simulations.  \bf Metallicities
greater than $\sim$0.1\,Z$_\odot$ are difficult to reconcile with the
observed G-dwarf distribution. \rm  \it Lower: \rm 
Present-day Milky Way radial oxygen abundance
profile predicted by our dual-infall phase chemical evolution model.
Observed abundances in HII 
regions are represented by symbols, while the curves
indicate model predictions for disk-phase infalling gas with primordial
(solid line), 20\% solar (dotted line) and 40\% solar (dashed line)
metallicity. All models assume primordial composition for the halo-forming
gas.  The three models satisfy the observed slope, although \bf
each increase in
metallicity shown here is accompanied by a 0.1\,dex increase in zero point.\rm
The mild overproduction of oxygen, even in the case of primordial infall, 
is a common aspect of Galactic chemical evolution models.
}
\end{figure}

The fact that most HVCs have metallicities of
$\sim$0.3\,Z$_\odot$ led Gibson et~al. (2001a) to conclude that not only
was this inconsistent with the Blitz et~al. (1999) and Braun \& Burton (1999)
pictures, but also with the classical Galactic
Fountain picture (in which metallicities near solar might be expected).
While we are still of that opinion, the conclusion should perhaps be 
tempered by the bottom panel of Figure~2.  \it If \rm HVCs
originate from a fountain at a Galactocentric distance
$\sim$5\,kpc (location
of the disk's present-day star formation maximum), metallicities
in the range $\sim$0.6$\rightarrow$1.3\,Z$_\odot$ would be 
expected (``solar'' being log(O/H)$+$12=8.9); 
\it if \rm HVCs originate near the solar circle 
($\sim$6$\rightarrow$12\,kpc), the range would extend to 
$\sim$0.2$\rightarrow$1.2\,Z$_\odot$.  These ranges are, technically,
upper limits, as dilution from near-pristine halo gas will push
the values down somewhat.  The point we wish to make here is that
the \it occasional \rm 
metallicity as low as $\sim$0.2\,Z$_\odot$ does not sound the
death-knell for the Galactic Fountain; that said, persistent values this
low are difficult to reconcile with a Galactic origin, unless dilution
by metal-poor halo gas is extremely efficient.

The next 12 months will see
Complex~WD (toward the Local Group's anti-barycentre) come under 
scrutiny,\footnote{Including our FUSE Cycle~2 GI program, in collaboration
with Mark Giroux.} lengthy revisits to Complex~C,\footnote{Including
several HST GI programs led by Bart Wakker and David Bowen, as well as
the FUSE Science Team's ongoing analyses.} the Mrk~205
HST/STIS analysis of Bowen et~al. (HST 
PID\#8625),\footnote{Mrk~205 intersects both
Complex~C and CHVC~125$+$41$-$207; the Bowen et~al. data should provide the
first useful metallicity determination for a CHVC.} and the first useful
upper limit on the metallicity of a CHVC (Sembach et~al. 2001).

\subsection{Kinematics}
As discussed in Gibson et~al. (2001a), arguments in support of an
intra-Local Group residency for HVCs, based upon the velocity distribution
(both its centroid and dispersion) being more ``favourable'' in the Galactic or
Local Group Standard of Rest (as opposed to the Local Standard of Rest) are
somewhat specious.  This is a necessary, but \it not \rm sufficient,
condition; \it any \rm model which results in the appropriate sinusoidal
behaviour in the $\ell$-$v_{\rm LSR}$ plane, which includes many
Galactic fountain and
Magellanic Cloud disruption scenarios, will result in a decreased 
velocity dispersion in the GSR and LGSR frames.

The number of catalogued HVCs below $\delta=+0^\circ$ has now grown
to $\sim$2000, with $\sim$10\% alone classified as CHVCs (Putman et~al.
2001).  It should be noted though, that
approximately half of the original Braun \& Burton (1999) CHVCs which lie
in the Putman et~al. 
overlap region have lost their ``compact'' status, and have 
been reclassified (due to the improved spatial resolution used by
Putman et~al., and
a stricter classification scheme).  Of futher interest, the 179 Putman et~al.
CHVCs are \it not \rm distributed randomly, but are in fact clustered into
three primary groups which lie within $\pm$25$^\circ$ of
the Galactic Plane, or near the South Galactic Pole.  The Putman et~al.
HVCs \it and \rm CHVCs
have nearly identical $v_{\rm GSR}$ and $v_{\rm LGSR}$ centroids.  Whether
that means both HVCs and CHVCs therefore have
the same origin remains unresolved.

We wish to comment upon the size-linewidth analysis of Combes \&
Charmandaris (2000).  These authors show that those HVCs which are
\it known \rm to reside in the Galactic halo, adhere to the 
molecular cloud size-linewidth relation.  Further, they note that the
Braun \& Burton
(1999) CHVCs would also follow this relation, \it if they had
typical distances of 20\,kpc\rm.  Invoking Occam's Razor, Combes \&
Charmandaris suggest therefore that most CHVCs are not of an
extragalactic nature.  While interesting, and possibly
correct, this conclusion should be tempered with the knowledge that dark matter
dominated Local Group objects (such as the aforementioned dSphs)
do not follow this self-same relationship.
Taken one step further, if CHVCs are also in fact dark matter dominated
entities (as both Blitz et~al. 1999 and Braun \& Burton 1999 argue), then
there would be no \it a priori \rm reason to force them to follow the
Galactic molecular cloud size-linewidth relation.

\subsection{Association with Extra-Local Group Systems}
Some of the more compelling arguments against the Blitz et~al. (1999) and
Braun \& Burton (1999) scenarios are due to Zwaan (2001) and
Charlton et~al. (2000).  Both teams surveyed nearby Local Group
analogs in an attempt to ascertain how common intragroup gas clouds are
elsewhere.  Charlton et~al. show that the statistics of
Mg{\small II} and Lyman limit absorbers in the spectra of background
QSOs are in disagreement with the extragalactic HVC scenarios.
Zwaan surveyed $\sim$4\,Mpc$^2$ of five nearby Local Group
analogs, down to an H{\small I} mass of 7$\times$10$^6$\,M$_\odot$
(4.5$\sigma$), also finding no intragroup clouds. \footnote{Counter-arguments 
have been presented by Blitz (2001).}  The proposed
HIPARK Survey at Parkes (led by Frank Briggs, with Martin 
Zwaan, David Barnes, and us) 
would have a 6$\sigma$ detection limit at 6\,Mpc of 
1$\times$10$^6$\,M$_\odot$, covering $\sim$200\,deg$^2$, and would 
offer more than an
order of magnitude improvement over any of the existing extra-Local Group
surveys.

An issue which has not received fair attention, but one which remains a
tantalising one nevertheless, is the observation that the incidence
HVC ``activity'' in isolated spirals is directly related to the 
underlying star formation rate (Schulman et~al. 1997).  
This certainly seems to be an important clue in the HVC ``mystery'', but
one which has received scant attention in the recent literature.

\subsection{Magellanic Stream}
The disruption of the Magellanic Clouds has long been recognised as a 
potentially crucial component of any complete HVC origin theory
(e.g. Wakker 1990; Ch.~5).  The discovery of the Leading Arm
Feature (LAF), counterpart to the trailing Magellanic Stream, demonstrates
that strong tidal forces are involved in disrupting our nearest
neighbours (Putman et~al. 1998).  Of further interest is the 
discovery that the Magellanic Stream is not confined to the canonical
linear H{\small I} stream.  As Gibson et~al. (2000) show,
gas associated with the Stream is seen in Mg{\small II}
along the III~Zw~2 sightline;
this has since been supplemented by the detection of the
Stream in Ly$\beta$ in FUSE observations of Mrk~335.  Coupled with NGC~7469,
we now have three clear detections of the Stream in the vicinity of the
MS~V concentration (Figure~3).  FUSE has observed many other sightlines
which project onto Figure~3 (and elsewhere in the vicnity of the Stream); 
examining these data for other Ly$\beta$
Stream ``signatures'' will be necessary in order to assess
the true extent of disrupted gas from the Magellanic Clouds.
We are currently
pursuing high-resolution N-body $+$ SPH simulations of the formation
and evolution of the Stream, LAF, and associated ``disrupted'' gas from the
Clouds.

\begin{figure}[ht]
\plotfiddle{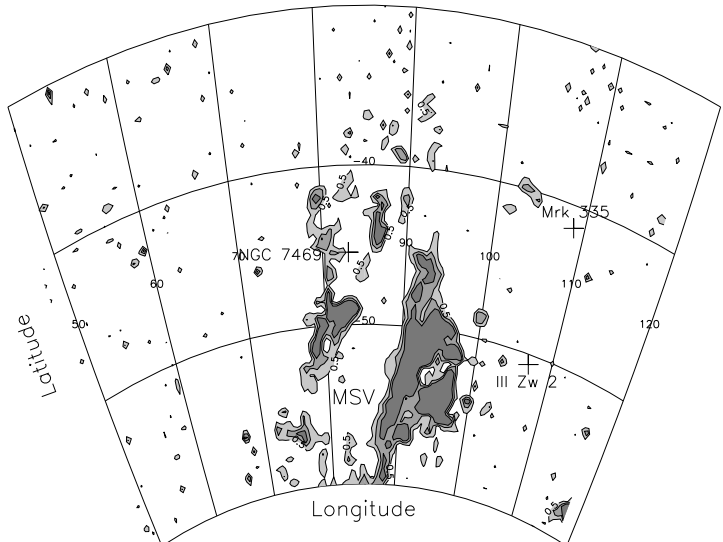}{2.60in}{0}{40}{40}{-130}{-15}
\caption{\footnotesize
Leiden-Dwingeloo Survey H{\small I} map of the MS~V concentration
within the Magellanic Stream.  Beyond the confines of the classical H{\small I}
filament, the Stream has been detected in ultraviolet absorption towards
NGC~7469, III~Zw~2, and Mrk~335 - more than 20\,kpc (projected)
from the canonical Stream!
}
\end{figure}

\subsection{Alternatives to $\Lambda$CDM}
When all else fails, should the HVCs and CHVCs not correspond to
$\Lambda$CDM halos, we may be faced with the prospect of 
venturing outside the confines of standard hierarchical clustering
scenarios.  Both Warm Dark Matter (WDM) and Self-Interacting Dark
Matter (SIDM) offer potentially viable mechanisms for suppressing
power at small scales.
That said, two of the best related analyses invoking WDM (Bode et~al. 2001)
and SIDM (Dav\'e et~al. 2001) show that low-mass halo suppression
is limited to factors of a few, as opposed to the
orders-of-magnitude required to reconcile theory with the Local Group
galaxy distribution.  Perhaps the solution retains
$\Lambda$CDM, but involves an efficient treatment of feedback, a l\`a
Chiu et~al. (2001).  The latter suppress low-mass gas halo
production with ``physics'', without suppressing the
number of dark halos.\footnote{This simulation was not evolved beyond 
redshift $z$=4, and so questions remain concerning the clustering
properties at redshift $z$=0.}

%\section{Summary}

\acknowledgments
We wish to thank Mary Putman,
Amiel Sternberg, Mark Giroux, Ken Sembach, Joss Bland-Hawthorn,
Dan Kelson, Josh Simon, and Bart Wakker, for many helpful correspondences.

\end{document}